\documentclass[useAMS,usenatbib]{mn2e}

\usepackage{txfonts,graphicx,amssymb,rotating}

\title[Disc photoevaporation II: evolutionary models]{Photoevaporation of protoplanetary discs II: evolutionary models and observable properties}

\author[R.D.~Alexander, C.J.~Clarke \& J.E.~Pringle]
  {R.D.~Alexander$^{1,2,}$\thanks{email: rda@jilau1.colorado.edu},
  C.J.~Clarke$^1$
  and J.E.~Pringle$^1$ \\
   $^1$ Institute of Astronomy, Madingley Road, Cambridge, CB3 0HA, UK\\
   $^2$ JILA, 440 UCB, University of Colorado, Boulder, CO 80309-0440, USA}

\begin{document}

\pagerange{\pageref{firstpage}--\pageref{lastpage}} \pubyear{2006}

\maketitle

\label{firstpage}

\begin{abstract}
We present a new model for protoplanetary disc evolution.  This model combines viscous evolution with photoevaporation of the disc, in a manner similar to \citet*{cc01}.  However in a companion paper \citep*{paper1} we have shown that at late times such models must consider the effect of stellar radiation directly incident on the inner disc edge, and here we model the observational implications of this process.  We find that the entire disc is dispersed on a time-scale of order $10^5$yr after a disc lifetime of a few Myr, consistent with observations of T Tauri (TT) stars.  We use a simple prescription to model the spectral energy distribution of the evolving disc, and demonstrate that the model is consistent with observational data across a wide range of wavelengths.  We note also that the model predicts a short ``inner hole'' phase in the evolution of all TT discs, and make predictions for future observations at mid-infrared and millimetre wavelengths.
\end{abstract}



\section{Introduction}\label{sec:intro}
For over twenty years the evolution, and eventual dispersal, of discs around young stars has been an important area of study, for theories of both star and planet formation.  It is now well-established that at an age of $\sim 10^6$yr most stars are surrounded by discs that are optically thick at optical and infrared wavelengths \citep{strom89,kh95}.  Observations at millimetre wavelengths show that these discs have masses that are typically a few percent of a solar mass \citep{beckwith90}, and so discs are widely believed to be potential sites for planet formation.  However at an age of $\sim 10^7$yr most stars are not seen to have discs, suggesting that disc lifetimes are typically a few Myr \citep[e.g.][]{haisch01}.  How stars lose their discs remains an unsolved question.

The distribution of T Tauri stars (TTs) at infrared wavelengths provides some insight into this problem.  These objects tend to fall into two distinct groups: those whose emission is consistent with a stellar photosphere plus an optically thick disc, and those which are compatible with purely photospheric emission.  [Note that these two classes usually coincide with the spectroscopic classifications of classical and weak-lined T Tauri stars (CTTs and WTTs) respectively.]  A number of authors have noted that very few transition objects are observed between the CTT and WTT loci \citep*{skrutskie90,kh95,act99,hartmann05}.  Observations at longer wavelengths show a similar behaviour, both in the mid-infrared \citep{persi00,bontemps01} and at millimetre wavelengths \citep{duvert00,aw05}.  These observations suggest that discs are dispersed very rapidly, with the dispersal time estimated to be $\sim 10^5$yr \citep{sp95,ww96}.  Moreover the simultaneous decline is disc emission across such a wide range in wavelength suggests that the dispersal is essentially simultaneous across the entire radial extent of the disc (see also discussion in \citealt*{tcl05}).

This ``two-time-scale'' behaviour is inconsistent with conventional models of disc evolution, such as viscous evolution models \citep{hcga98} or models of magnetospheric clearing \citep{act99}.  Such models predict power-law declines in disc properties, and therefore predict dispersal times which are always of the same order as the disc lifetime.  However \citep*[][hereafter CGS01]{cc01} showed that models which combine photoevaporation of the disc with viscous evolution can reproduce this two-time-scale behaviour.  In this model, known as the ``UV-switch'' model (CGS01), ionizing radiation from the central star produces a photoevaporative wind at large radii.  Detailed models of photoevaporative winds were constructed by \citet[][see also \citealt*{hollppiv}]{holl94} for the cases of both weak and strong stellar winds.  In the case of TTs we consider only the weak stellar wind case.  In this case ionizing radiation from the star creates an ionized layer on the surface of the disc, with conditions akin to an H\,{\sc ii} region.  Beyond some critical radius, known as the gravitational radius, the local thermal energy of the ionized is greater than its gravitational energy and the gas escapes as a wind.  The gravitational radius is therefore given by
\begin{equation}\label{eq:R_g}
R_{\mathrm g} = \frac{GM_*}{c_{\mathrm s}^2} = 8.9 \left(\frac{M_*}{\mathrm M_{\odot}}\right) \mathrm {AU}
\end{equation}
where $c_{\mathrm s}$ is the sound speed of the ionized gas, typically 10kms$^{-1}$.  \citet{holl94} show that the wind rate is determined by the the density at the ionization front, and find an integrated mass-loss rate of
\begin{equation}\label{eq:wind_rate}
\dot{M}_{\mathrm {wind}} \simeq 4.4\times 10^{-10} \left(\frac{\Phi}{10^{41}\mathrm s^{-1}}\right)^{1/2} \left(\frac{M_*}{1\mathrm M_{\odot}}\right)^{1/2} \mathrm M_{\odot} \mathrm {yr}^{-1},
\end{equation}
where $\Phi$ is the ionizing flux produced by the star.  More recent studies have extended the study of the details in a number of ways \citep*{ry97,jhb98,font04}, and recent hydrodynamic modelling has resulted in slight modification of the quantitative results.  When hydrodynamic effects are considered the ``effective $R_{\mathrm g}$'' is reduced by a factor of 5 \citep{liffman03,font04}, and the mass-loss rate is reduced by a factor of around 3 \citep{font04}.  However the qualitative behaviour is unchanged from that of \citet{holl94}.

The so-called ``UV-switch'' model of CGS01 couples a photoevaporative wind to a simple disc evolution model.  At early times in the evolution the accretion rate through the disc is much larger than the wind rate, and the wind has a negligible effect.  However at late times photoevaporation becomes important, depriving the disc of resupply inside $R_{\mathrm g}$.  At this point the inner disc drains on its own, short, viscous time-scale, giving a dispersal time much shorter than the disc lifetime.  A number of similar studies have now been conducted \citep*{mjh03,acp03,ruden04,tcl05}, and this class of models show a number of attractive properties.

However CGS01 highlighted two key problems with the model.  Firstly, the model requires that TTs produce a rather large ionizing flux, of order $10^{41}$ ionizing photons per second.  They also found that the outer disc, beyond $R_{\mathrm g}$ was dispersed much too slowly to satisfy millimetre observations of WTTs, a finding re-affirmed by recent sub-millimetre observations \citep{aw05}.  We have previously shown that it is reasonable to treat TT chromospheres as having a constant ionizing flux in the range $\sim 10^{41}$--$10^{44}$s$^{-1}$ \citep*{chrom}, and we now seek to address the ``outer disc problem''.  In a companion paper \citep*[][ hereafter Paper I]{paper1}, we highlighted an important flaw in the UV-switch model.  The UV-switch model relies on the wind parametrization of \citet{holl94}, which assumes that the disc is extremely optically thick to Lyman continuum photons at all radii.  \citet{holl94} find that the diffuse (recombination) field dominates the photoevaporation at all radii of interest, as the direct field suffers extremely strong attenuation by the disc atmosphere.  However we note that at late times in the UV-switch model the inner disc is drained, and is therefore optically thin to ionizing radiation.  Consequently the direct field is important after the inner disc has drained.  In Paper I we constructed detailed hydrodynamic models of the wind driven by the direct field, and derived a functional form for the mass-loss rate:
\begin{eqnarray}\label{eq:mout_anal}
\dot{M}(<R_{\mathrm {out}}) = 1.73\times10^{-9} \, \frac{CD}{a-2} \, \mu \, \left(\frac{\Phi}{10^{41}\mathrm s^{-1}}\right)^{1/2} \left(\frac{H/R}{0.05}\right)^{-1/2} 
\nonumber \\
\times \left(\frac{R_{\mathrm {in}}}{3\mathrm{AU}}\right)^{1/2}  \left[1 - \left(\frac{R_{\mathrm {in}}}{R_{\mathrm {out}}}\right)^{a-2}\right] \, \mathrm M_{\odot}\mathrm{yr}^{-1} \, .
\end{eqnarray}
Here $C$ and $D$ are order-of-unity scaling constants, $\mu$ is the mean molecular weight of the gas, $a$ is a power-law index, $H/R$ is the ratio of the disc scale-height to radius, and $R_{\mathrm {in}}$ and $R_{\mathrm {out}}$ are the inner and outer disc radii respectively.  Our numerical analysis fixed the values of the scaling constants to be $a=2.42\pm0.09$ and $(CD)=0.235\pm0.02$ (for $H/R=0.05$).  As noted in Paper I, both the geometry of the radiative transfer problem and the form of the wind are qualitatively similar to the strong wind case of \citet{holl94}.  However the effect of the direct radiation field is to increase the efficiency of the wind.  Consequently the mass-loss rate due to direct photoevaporation is around an order of magnitude larger than that from the diffuse field, and is significant at late stages of the evolution.

The diffusion equation for the evolution of disc surface density $\Sigma(R,t)$ is \citep{lbp74,pringle81,acp03}
\begin{equation}\label{eq:1ddiff}
\frac{\partial \Sigma}{\partial t} = \frac{3}{R}\frac{\partial}{\partial R}\left[ R^{1/2} \frac{\partial}{\partial R}\left(\nu \Sigma R^{1/2}\right) \right] - \dot{\Sigma}_{\mathrm {wind}}(R,t) \, ,
\end{equation}
where $\nu$ is the kinematic viscosity and the term $\dot{\Sigma}_{\mathrm {wind}}(R,t)$ represents the mass-loss due to photoevaporation.  CGS01 solved this equation using the ``weak-wind'' profile of \citet{holl94}.  At some point in the evolution the mass-loss rate from the wind falls to a level comparable to the accretion rate through the disc, and at this point the disc is rapidly drained inside $R_{\mathrm g}$.  However, as mentioned above, CGS01 neglect the influence of the direct radiation field after this inner draining occurs.  Consequently they find that the time-scale for dispersal of the outer disc is limited by the time material takes to diffuse inward to $R_{\mathrm g}$ (as the most of the mass-loss occurs close to $R_{\mathrm g}$).  Further, the $R^{-5/2}$ dependence of the wind profile at large radii means that the mass-loss rate due to photoevaporation decreases significantly with time as the inner edge of the disc moves outward.  Consequently the dispersal of the outer disc occurs on the viscous time-scale of the {\it outer} disc, and thus the outer disc is dispersed in a time comparable to the disc lifetime, much too slowly to satisfy observational constraints.  We suggest that photoevaporation by the direct radiation field results in a dispersal time significantly shorter than that predicted by CGS01, and now seek to model the effects of this process on the evolution of the outer disc.

In this paper we seek to incorporate the result of Paper I into models of disc evolution.  We do this by solving the equation for the evolution of the surface density of a geometrically thin disc, including the photoevaporative wind as a sink term.  In Section \ref{sec:timescales} we present a simple time-scale analysis, which demonstrates the significance of the direct radiation field.  In Section \ref{sec:disc_model} we construct a numerical model of disc evolution in the presence of a photoevaporative wind.  We the use a simple prescription to model the observed spectral energy distribution (SED) of the evolving disc, and construct a set of models which cover a broad range in parameter space (Section \ref{sec:SED_model}).  In Section \ref{sec:results} we present our results, comparing the predicted SEDs to recent observational data.  In Section \ref{sec:dis} we discuss the implications and limitations of our results, and in Section \ref{sec:summary} we summarize our conclusions.

\section{Evolutionary time-scales}\label{sec:timescales}
It is useful at this point to consider the evolutionary time-scales predicted by the different models.  In the original model of CGS01, the inner disc drains on approximately the viscous time-scale at the draining radius, $t_{\nu}(R_{\mathrm g})$, which is significantly shorter than the disc lifetime to that point.  However the outer disc drains on a much longer time-scale, comparable to the disc lifetime.  After the inner disc has drained there is no accretion on to the star, so the only mass-loss is due to the wind.  Consequently, if we neglect viscous evolution during clearing, the time-scale for the wind to clear the disc out to a radius $R>R_{\mathrm g}$, $t_{\mathrm c}(R)$, is given by
\begin{equation}\label{eq:tc}
t_{\mathrm c}(R) = \frac{M_{\mathrm d}(<R)}{\dot{M}_{\mathrm {wind}}(R)} \, ,
\end{equation}
where $M_{\mathrm d}(<R)$ is the total disc mass at radii $<R$ and $\dot{M}_{\mathrm {wind}}(R)$ is the total mass-loss rate from the wind, integrated from radius $R$ outwards.
Neglecting numerical factors of order unity, the viscous time-scale for the evolution of an accretion disc is given by
\begin{equation}\label{eq:tnu}
t_{\nu}(R) = \frac{M_{\mathrm d}(<R)}{\dot{M}_{\mathrm d}} \, ,
\end{equation}
where $\dot{M}_{\mathrm d}$ is the disc accretion rate.  We can therefore combine Equations \ref{eq:tc} \& \ref{eq:tnu} and express the clearing time as
\begin{equation}
t_{\mathrm c}(R) = t_{\nu}(R)\frac{\dot{M}_{\mathrm d}}{\dot{M}_{\mathrm {wind}}(R)} \, .
\end{equation}
Clearing occurs when the disc accretion rate $\dot{M}_{\mathrm d}$ falls to a level comparable to the diffuse wind rate (which is constant, see Equation \ref{eq:wind_rate}), and therefore the clearing time-scale depends only on $\dot{M}_{\mathrm {wind}}(R)$.  If $\dot{M} _{\mathrm {wind}}(R)$ is a decreasing function of $R$ then accretion will dominate over the wind as the disc evolves, whereas if $\dot{M}_{\mathrm {wind}}(R)$ increases with $R$ then the wind will dominate the evolution.

In the model of CGS01, where the diffuse field mass-loss profile is adopted throughout, the wind profile takes the form $\dot{\Sigma}_{\mathrm {wind}}(R) \propto (R/R_{\mathrm g})^{-5/2}$.  Consequently the total wind mass-loss rate, integrated from an inner edge radius $R$ outwards, is $\dot{M}_{\mathrm {wind}}(R) \propto R^{-1/2}$, and we can normalise by noting that $t_{\mathrm c}(R_{\mathrm g}) = t_{\nu}(R_{\mathrm g})$.  Therefore the clearing time-scale is given by
\begin{equation}
t_{\mathrm c}(R) = t_{\nu}(R) \left(\frac{R}{R_{\mathrm g}}\right)^{1/2} \, .
\end{equation}
Thus we see that in the UV-switch model (CGS01) the time-scale to clear the disc out to a radius $R$ is longer than the viscous time-scale for all $R>R_{\mathrm g}$.  Consequently in this model viscosity dominates: mass is lost only after it has had time to diffuse inwards towards the inner disc edge.  Thus while the inner disc satisfies the two-time-scale behaviour demanded by observations (see Section \ref{sec:intro}), the outer disc is dispersed much too slowly to satisfy the data.

However if we consider direct photoevaporation of the outer disc we see a very different behaviour.  In this case, as seen in Equation \ref{eq:mout_anal} (see also Paper I), the integrated mass-loss from a disc with an inner edge at radius $R$ is $\dot{M}_{\mathrm {wind}} \propto R^{1/2}$.  Consequently in this case the clearing time-scale is given by
\begin{equation}
t_{\mathrm c}(R) = t_{\nu}(R) \left(\frac{R}{R_{\mathrm g}}\right)^{-1/2} \, .
\end{equation}
Thus we see that in the case of direct photoevaporation viscosity becomes progressively {\it less} significant as the inner edge evolves outwards, and that the evolution is instead dominated by the wind.  Consequently we predict a much faster dispersal of the outer disc than that originally predicted by CGS01. For TT parameters the draining radius is of order AU, and TT discs are observed to be up to several hundred AU in size.  Thus we predict a clearing time for the entire disc that is $\sim$1--10\% of the viscous scaling time.  Such behaviour satisfies the two-time-scale constraint across the entire radial extent of the disc.


\section{Disc model}\label{sec:disc_model}
We solve the diffusion equation for the disc surface density (Equation \ref{eq:1ddiff}) using a standard first-order explicit scheme \citep*[e.g.][]{pvw86}, with a grid of points equispaced in $R^{1/2}$.  Following \citet{hcga98} and CGS01, we adopt a kinematic viscosity $\nu$ which scales linearly with radius, so
\begin{equation}
\nu(R) = \nu_0 \frac{R}{R_0}
\end{equation}
for some scaling value $\nu_0(R_0)$ at a scale radius $R_0$.  Such a viscosity law is consistent with an $\alpha$-prescription \citep{ss73} if the disc temperature at the midplane scales as $R^{-1/2}$, and arguments in favour of this viscosity law are discussed in \citet{hcga98}.  We adopt an initial surface density profile consistent with the similarity solution of the diffusion equation \citep{lbp74,hcga98}.  Again following CGS01, we define this profile to have the form 
\begin{equation}
\Sigma(R)=\frac{M_{\mathrm d}(0)}{2\pi R_0 R}\exp(-R/R_0) \, ,
\end{equation}
for an initial disc mass $M_{\mathrm d}(0)$.  In this form $1/\mathrm e$ of the disc mass is initially at $R>R_0$, with an exponential decline in surface density at radii beyond $R_0$.  However in practice the results are not sensitive to the form of the initial profile, and this form is chosen primarily as a simple means of parametrizing the initial disc mass.  With this viscosity law the viscous scaling time, which governs the evolution of the disc, is given by
\begin{equation}\label{eq:visc_time}
t_{\nu} = \frac{R_0^2}{3\nu_0} \, ,
\end{equation}
and consequently the initial accretion rate at the origin is given by
\begin{equation}\label{eq:mdot_visc}
\dot{M}_{\mathrm d}(R=0,t=0) = \frac{3 M_{\mathrm d}(0) \nu_0}{2R_0^2} \, .
\end{equation} 
This fixes the scaling constant $\nu_0$, and thus the disc model is entirely specified by the three parameters $M_{\mathrm d}(0)$, $\dot{M}_{\mathrm d}(0,0)$ and $R_0$.  Initially we adopt $M_{\mathrm d}(0)=0.05$M$_{\odot}$, $\dot{M}_{\mathrm d}(0,0) = 5.0\times 10^{-7}$M$_{\odot}$yr$^{-1}$ and $R_0=10$AU.  We use 1000 grid points, equispaced in $R^{1/2}$, which span the radial range $[0.0025\mathrm{AU},2500\mathrm{AU}]$. We also adopt adopt zero-torque boundary conditions throughout (i.e.~we set $\Sigma = 0$ at the grid boundaries) but note that the spatial domain is always large enough that the outer boundary condition has no effect on the results.  

At early times in the evolution we use the wind profile of \citet[][ kindly provided in numerical form by Ian McCarthy]{font04}, which incorporates more detailed hydrodynamics than the original work of \citet{holl94}.  However once the inner disc has been drained we alter the mass-loss profile to reflect the influence of direct photoevaporation on the outer disc.  Thus it is necessary to define a numerical criterion for ``draining'', in order to indicate when to change to the direct profile.  Additionally, as the form of the direct profile is normalised at the inner disc edge it is necessary to define the inner edge numerically also.  

By inspection of the mass-loss profiles we see that in both the diffuse and direct cases the mass-loss rate scales as
\begin{equation}
\dot{M} \propto \Phi^{1/2} \, .
\end{equation}
Numerical analysis of the scaling constants shows that for equal ionizing fluxes the direct mass-loss rate exceeds that due to the diffuse field by a factor of $8.8$ (for $M_*=1$M$_{\odot}$).  Thus if the direct flux reaching the disc is greater than approximately 0.01 of its total value the direct wind exceeds the diffuse wind.  Consequently our criterion for the ``transition'' between the two wind profiles is that the optical depth to ionizing photons, $\tau$, along the disc midplane to the draining radius ($0.2R_{\mathrm g}$) satisfies
\begin{equation}
\exp(-\tau)=0.01 \qquad \Rightarrow \qquad \tau=4.61 \, .
\end{equation}
We evaluate the optical depth as $\tau=N\sigma_{13.6{\mathrm eV}}$, where $N$ is the column density along the disc midplane.  Additionally, once the inner disc has drained we define the inner edge $R_{\mathrm {in}}$ to be the radius at which the optical depth reaches this critical value.  After draining we adopt the mass-loss profile derived in Paper I (for $H/R=0.05$):
\begin{equation}
\dot{\Sigma}_{\mathrm {wind}}(R) = 2 CD \mu m_{\mathrm H} c_{\mathrm s}  n_{\mathrm {in}} \left(\frac{R}{R_{\mathrm {in}}}\right)^{-a}\, ,
\end{equation}
where
\begin{equation}
n_{\mathrm {in}} = \left(\frac{\Phi}{4\pi \alpha_B \frac{H}{R} R_{\mathrm {in}}^3}\right)^{1/2} \, .
\end{equation}
Here $\alpha_B$ is the Case B recombination coefficient for atomic hydrogen at $10^4$K, which has a value of $\alpha_B=2.6\times10^{-13}$cm$^3$s$^{-1}$ \citep{allen}.  We adopt the best-fitting scaling constants from Paper I of $(CD)=0.235$ and $a=2.42$, and set $\mu=1.35$ (following \citealt{holl94}, CGS01).

\subsection{Results}\label{sec:ref_results}
\begin{figure}
\centering
        \resizebox{\hsize}{!}{
        \includegraphics[angle=270]{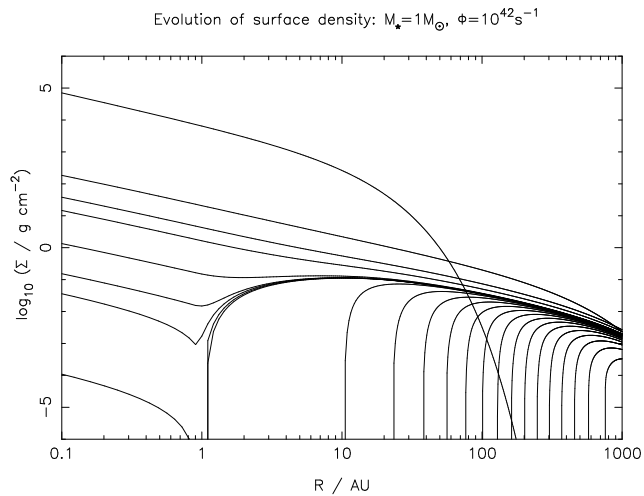}
        }
        \caption{Evolution of surface density in model which incorporates direct photoevaporation.  Snapshots of the surface density are plotted at $t=0$, 2.0, 4.0, 5.9, 6.0, 6.01, 6.02, 6.02\ldots6.18Myr.  At $t=6.20$Myr the surface density is zero across the entire grid.  After the inner disc is drained direct photoevaporation disperses the outer disc very rapidly.}
        \label{fig:uv_switch_plus}
\end{figure}
Fig.\ref{fig:uv_switch_plus} shows the evolution of the surface density for the reference model.  This model has $M_*=1$M$_{\odot}$, $M_{\mathrm d}(0)=0.05$M$_{\odot}$, $\dot{M}_{\mathrm d}(0,0) = 5.0\times 10^{-7}$M$_{\odot}$yr$^{-1}$ and $R_0=10$AU.  The results shown are for $\Phi=10^{42}$s$^{-1}$.  The inner disc begins to drain at $t=6.01$Myr, and the disc is cleared to the outer grid radius (2500AU) at $t=6.20$Myr.  At the point where the inner disc begins to drain the total disc mass remaining is approximately 0.002M$_{\odot}$, or approximately 2M$_{\mathrm {Jup}}$.  

In order to investigate the dependence on the value of $H/R$ the same disc model was run with $H/R=0.1$ (and therefore best-fitting parameters of $(CD)=0.60$ and $a=4.50$, see Paper I).  In this case the inner disc draining is identical to the case of $H/R=0.05$, as the early evolution depends only on the diffuse wind and is independent of $H/R$.  The outer disc is cleared somewhat more slowly by the direct wind than for $H/R=0.05$, with the outer grid radius not reached until $t=6.27$Myr.  However this represents a change of less than a factor of two in the time required to clear the outer disc, so we are satisfied that the choice of $H/R$ is not a significant factor in evolution of the outer disc.

A second version of this model was run with $\nu \propto R^{3/2}$, in order to investigate the effect of varying the viscosity law.  The numerical scaling constants in the similarity solution vary with the viscosity law \citep{lbp74,hcga98}, but while the details differ the qualitative behaviour of the model is unchanged.  The steeper surface density profile that results from this new viscosity law means that the inner disc draining is somewhat slower than in the reference model, but this is countered by a more rapid clearing of the outer disc (as a smaller fraction of the disc mass now resides at large radii).  The disc is still dispersed on a time-scale some 1--2 orders of magnitude shorter than the disc lifetime\footnote{Here the term ``lifetime'' refers to the age of the disc, $t$, when the disc is dispersed.}, and the only significant consequence of changing the viscosity law is the expected modification of the surface density profile.

Thus, as predicted in Section \ref{sec:timescales}, when direct photoevaporation is taken into account the entire disc is dispersed on a time-scale approximately 2 orders of magnitude shorter than the disc lifetime.  Thus it seems that this model has solved the ``outer disc problem'' that affected the original UV-switch model of CGS01.  In order to compare to observed data, however, it is necessary to model the observable properties of the disc.


\section{Observable consequences: spectral energy distributions}\label{sec:SED_model}
In order to compare the results of the model to observed data it is necessary to use the model to generate spectral energy distributions (henceforth SEDs).  Following \citet{hcga98} and CGS01, we assume that the disc is vertically isothermal and emits as a blackbody.  Therefore the flux emitted by the disc at frequency $\nu$ is\footnote{Note that we have now used the symbol $\nu$ to denote both frequency and kinematic viscosity. $\nu$ denotes frequency only in Equations \ref{eq:nu1}--\ref{eq:nu2}, and it should be obvious from context to which quantity the symbol refers.}
\begin{equation}\label{eq:nu1}
F_{\nu} = \frac{\cos i}{4\pi d^2} \int 2\pi R B_{\nu}\left(T(R)\right) \left[1-\exp\left(-\frac{\tau_{\nu}}{\cos i}\right)\right] dR \, ,
\end{equation}
where $d$ is the Earth-star distance (taken to be 140pc, the distance of the Taurus-Auriga cloud), $B_{\nu}(T)$ is the Planck function and $i$ is the inclination angle of the disc ($i=0$ is face-on).  $T(R)$ is the radial temperature profile of the disc, and the optical depth $\tau_{\nu}$ is evaluated as
\begin{equation}
\tau_{\nu} = \kappa_{\nu}\Sigma(R) \, .
\end{equation}
We adopt a standard power-law for the dust opacity $\kappa_{\nu}$ \citep{beckwith90}:
\begin{equation}\label{eq:nu2}
\kappa_{\nu} = 0.1\frac{\nu}{10^{12}\mathrm {Hz}} \quad  \mathrm {cm}^2 \, \mathrm g^{-1} \, .
\end{equation}
We add the stellar contribution to the SED as the total flux emitted by a blackbody of temperature $T_*$ and radius $R_*$. The values of these two parameters are taken from the pre-main sequence models of \citet[][ kindly provided in electronic form by Chris Tout]{tlb99}.  TT stars are contracting due to gravitational collapse, so for a given stellar mass both $T_*$ and $R_*$ vary somewhat with age. We adopt the median age of the Taurus-Auriga cloud, 2Myr \citep{ps00,hartmann01}, throughout.  

\begin{figure}
\centering
        \resizebox{\hsize}{!}{
        \includegraphics[angle=270]{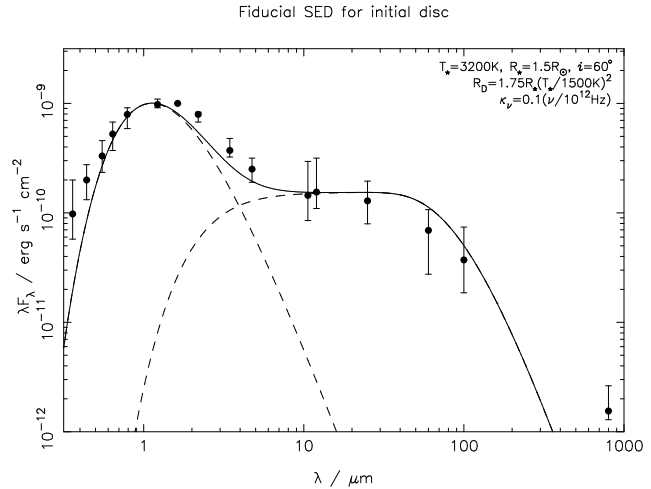}
        }
        \caption{SED produced by our model at $t=0$ for $T_*=3200$K and $R_*=1.5$R$_{\odot}$, values consistent with a pre-main-sequence star of mass 0.3--0.4M$_{\odot}$.  The SED produced by the model is shown as a solid line, with the individual contributions from the stellar black-body and the disc shown as dashed lines.  The points and error bars are the ``median SED'' of CTTs in Taurus-Auriga, taken from \citet{dalessio99}.  The two points at the shortest wavelengths are for $U$ and $B$ band data, and show the UV excess typical of CTTs.}
        \label{fig:median_SED}
\end{figure}

Obviously the disc temperature adopted is crucial to the resultant SED.  Further, we wish to investigate the effect of varying the stellar mass in our models, so it is necessary to define $T(R)$ in such a way that it only depends on the stellar mass $M_*$.  We adopt a ``flared reprocessing disc'' \citep{kh87} power-law profile:
\begin{equation}
T(R) = T_{\mathrm D} \left(\frac{R}{R_{\mathrm D}}\right)^{-1/2} \qquad , \qquad R \ge R_{\mathrm D}
\end{equation}
where the normalisation condition is set by the dust destruction radius $R_{\mathrm D}$.  At radii smaller than this the temperature is set to zero, as no dust can survive here and we assume that the opacity due to the gas alone is negligible.  We adopt a dust destruction temperature of $T_{\mathrm D}=1500$K.   Additionally, we adopt a minimum disc temperature of 10K to account for external heating of the disc (e.g.~cosmic rays or diffuse UV, see \citealt{hcga98}).  By assuming that the stellar irradiation balances the emission from the vertical disc edge at $R_{\mathrm D}$ we find that
\begin{equation}\label{eq:temp_norm}
R_{\mathrm D} = A R_* \left(\frac{T_*}{T_{\mathrm D}}\right)^2 \, ,
\end{equation}
where $A$ is a constant of proportionality that reflects how efficiently the disc radiates.  100\% efficiency (the ``small grains approximation'') gives $A=\sqrt{2}$, but by comparison to the observed ``median SED'' for Taurus-Auriga \citep[][ see Fig.\ref{fig:median_SED} below]{dalessio99} we find that $A=1.75$ provides a better fit to the data.  In adopting a value of $A>\sqrt{2}$ we essentially assume that the disc edge at $R_{\mathrm D}$ does not radiate as a blackbody (i.e.~it radiates with less than 100\% efficiency), an assumption verified by detailed radiative transfer models \citep[e.g.][]{dalessio05b}.  We note at this point that the disc temperature at the ``draining radius'' of $\simeq0.2R_{\mathrm g}$ does not vary significantly with stellar mass.  Obviously $R_{\mathrm g}$ scales linearly with $M_*$, but the effect of this on the disc temperature is offset by the variation of $T_*$ with stellar mass.  We find that $T(0.2R_{\mathrm g})=540$K for $M=0.2$M$_{\odot}$, but is only slightly smaller (410K) for $M=2.0$M$_{\odot}$.  Consequently, while the draining radius scales linearly with stellar mass, the colour change resulting from inner disc draining is not especially sensitive to stellar mass.

In addition to evaluating the SEDs, we use the predicted SEDs to generate magnitudes in the various photometric bands of the {\it 2-Micron All Sky Survey} (henceforth 2MASS) and the {\it Spitzer Space Telescope}.  We use the filter transmission functions given on the 2MASS and {\it Spitzer Science Center} websites\footnote{{\tt http://www.ipac.caltech.edu/2mass/} and {\tt http://ssc.spitzer.caltech.edu/} respectively.}, and use the given zero-point fluxes to convert these fluxes to magnitudes.

In order to test this model we compare the predicted SED with the ``median SED'' of CTTs in Taurus-Auriga \citep{dalessio99}.  We adopt stellar parameters of $T_*=3200$K and $R_*=1.5$R$_{\odot}$, consistent with $M_* =0.3$--0.4M$_{\odot}$ (depending on age), and use the surface density profile of the reference disc at $t=0$.  As seen in Fig.\ref{fig:median_SED}, the model reproduces the observed data well out to a wavelength of $\simeq 100$$\mu$m, but rather under-predicts the flux at millimetre wavelengths.  This apparent error is not of great concern, however, as in the initial disc configuration mass is mostly confined to small radii, with the result that the disc is optically thick at millimetre wavelengths.  At later times the disc expands and becomes optically thin, boosting the millimetre flux.  Additionally, once the disc is optically thin the emitted flux is very sensitive to the total disc mass, and our reference model has rather a low initial disc mass (0.05M$_{\odot}$).  More massive discs at later times (once the disc has spread beyond the scale radius $R_0$) match observed millimetre fluxes much better. In the {\it Spitzer} IRAC bands, our model gives colours of $[3.6]-[4.5]=0.46$, $[4.5]-[5.8]=0.57$, and $[5.8]-[8.0]=0.91$.  These compare favourably with the values for the median SED ($[3.6]-[4.5]=0.40$, $[4.5]-[5.8]=0.52$, and $[5.8]-[8.0]=0.83$, \citealt{hartmann05}), suggesting that the model predicts IRAC colours to an accuracy of around $\pm0.1$ mag.

\subsection{Model set}
In order to study the behaviour of the SED a series of disc models were run with different stellar masses.  As seen above, the outer disc evolution is not especially sensitive to the value of $H/R$, so when considering direct photoevaporation we adopt the best-fitting wind profile for $H/R=0.05$.  There are therefore five free parameters in the disc model: stellar mass $M_*$, initial disc mass $M_{\mathrm d}(0)$, initial accretion rate $\dot{M}_{\mathrm d}(0,0)$, scaling radius $R_0$ and ionizing flux $\Phi$.  Therefore it is necessary to evaluate these parameters self-consistently in order to study the effect of a single parameter.  In order to achieve this we assume that both the initial disc mass and the disc scaling radius scale linearly with $M_*$, and we adopt the normalisation conditions
\begin{equation}
R_0 = 10{\mathrm {AU}} \, \frac{M_*}{1\mathrm M_{\odot}} \, ,
\end{equation}
and 
\begin{equation}
M_{\mathrm d}(0) = 0.15M_* \, .
\end{equation}
The initial accretion rate is dependent on the disc viscosity.  As noted above, we adopt a viscosity law of the form $\nu \propto R$.  We now fix the normalisation by assuming that that the viscosity parameter $\alpha$ \citep{ss73} and disc aspect ratio $H/R$ are independent of $M_*$.  The orbital and viscous time-scales at $R_0$ are related by
\begin{equation}\label{eq:tnu_torb}
\frac{t_{\nu}}{t_{\mathrm {orb}}} \simeq \frac{1}{\alpha}\left(\frac{R}{H}\right)^2 \, .
\end{equation}
The relationships between the accretion rate, viscosity and viscous time-scale were given in Equations \ref{eq:visc_time} \& \ref{eq:mdot_visc}.  We combine these expressions to fix the initial accretion rate as
\begin{equation}
\dot{M}_{\mathrm d}(0,0) = \frac{M_{\mathrm d}(0)}{2t_{\nu}} \, .
\end{equation}
For typical parameters Equation \ref{eq:tnu_torb} gives $t_{\nu} \simeq 1000 t_{\mathrm {orb}}$, where $t_{\mathrm {orb}}$ is simply the Keplerian orbital time at $R_0$.  (This gives $t_{\nu}=3.2\times 10^4$yr at $R_0=10$AU.)  Thus the disc parameters are specified in a manner which depends only on the stellar mass $M_*$, due to the manner in which $M_{\mathrm d}(0)$ and $R_0$ are specified.  This fiducial disc model was evaluated for stellar masses of $M_* = 0.2$, 0.5, 1.0 \& 2.0M$_{\odot}$, and for each model the SED was evaluated as a function of time at an inclination angle of $i=60^{\circ}$ (i.e.~$\cos i=0.5$, the mean inclination of a random sample).  We set $\Phi = 10^{42}$s$^{-1}$ in all of these models.  In order to explore parameter space further four additional disc models were run for $M_* = 1.0$M$_{\odot}$, with $R_0=5$AU, $\Phi=10^{43}$s$^{-1}$, $M_{\mathrm d}(0) = 0.15M_*$ and $t_{\nu} = 5000 t_{\mathrm {orb}}$.  Lastly the SED for the fiducial 1M$_{\odot}$ model was also evaluated for an inclination angles of $i=0$ (i.e.~a face-on disc) and $i=80^{\circ}$ (i.e.~a nearly edge-on disc).  Thus one group of models explores the effect of stellar mass on the observed disc emission, while the second group explores the consequences of varying the disc parameters for a fixed stellar mass.

Magnitudes were evaluated in the $J$, $H$ and $K_{\mathrm s}$ 2MASS bands, and the four IRAC bands (which have central wavelengths of 3.6, 4.5, 5.8 \& 8.0$\mu$m respectively).  In addition magnitudes were calculated in the {\it Spitzer} 24$\mu$m MIPS band\footnote{Note that the zero-point flux in the 24$\mu$m MIPS band, 7.3Jy, is rather uncertain.  Consequently the absolute value of the magnitudes in this band are subject to systematic errors.  However this merely shifts the zero-point of the magnitude system, and any trends are unaffected.}.  The flux at $850$$\mu$m was also measured, for comparison to sub-millimetre (SCUBA) observations.  Each of these four wavebands (observed by 2MASS, IRAC, MIPS and SCUBA) probes a different region of the disc, with each providing different constraints.  The 2MASS bands primarily observe the stellar flux, and therefore provide a valuable normalisation condition.  The IRAC bands probe the inner disc, at the draining radius and smaller radii, while the 24$\mu$m MIPS band probes the emission at somewhat larger radii, beyond the initial draining radius.  The emission in both the IRAC and MIPS bands is (mostly) optically thick (the longer wavelength bands show weak optical depth effects), and so is sensitive only to the disc temperature and to whether or not the disc has drained.  Therefore small changes in the disc model do not have a significant effect here.  Lastly the flux at 850$\mu$m, as measured by SCUBA, measures optically thin emission from the entire disc, and is therefore rather sensitive to a number of the parameters in the disc model.  

We now seek to compare the results from our disc models to observed data.  In order to do this we have created a composite dataset, using observations taken from the literature.  We use data from recent surveys of the Taurus-Auriga cloud, from the IRAC observations of \citet{hartmann05} and the SCUBA observations of \citet{aw05}.  (\citealt{hartmann05} also list 2MASS magnitudes for all sources.)  We include only the sources which are unambiguously included in both samples, rejecting any binaries which are resolved by IRAC but not by SCUBA.  This leaves a total of 42 objects: 29 CTTs, 12 WTTs, and the possible transition object CoKu Tau/4 \citep[see][]{forrest04,dalessio05}.  All but 3 of the CTTs are detected by SCUBA, with upper limits only found for DP Tau, CIDA 11 and CIDA 12.  (These latter two objects are close to the brown dwarf limit and may not be ``true'' TTs.)  By contrast only 1 of the WTTs (LkH$\alpha$332 G1) is detected at 850$\mu$m, confirming that, in general, disc dispersal occurs simultaneously over the entire radial extent of the disc \citep[see discussion in][]{aw05}.


\section{Results}\label{sec:results}
\begin{table}
 \centering
  \begin{tabular}{|ccccccc|}
  \hline
  $M_*$ & $\Phi$ & $M_{\mathrm d}(0)$ & $R_0$ & $t_{\nu}/t_{\mathrm {orb}}$ & $t_1$ & $t_2$ \\
  M$_{\odot}$ & $10^{42}$s$^{-1}$ & $M_*$ & AU & at $R=R_0$ & Myr & Myr \\
  \hline
1.0 & 1.0 & 0.15 & 10.0 & 1000 & 8.37 &  8.47 \\
0.2 & 1.0 & 0.15 & 2.0 & 1000 & 3.52 & 3.57  \\
0.5 & 1.0 & 0.15 & 5.0 & 1000 & 6.13 & 6.21  \\
2.0 & 1.0 & 0.15 & 20.0 & 1000 & 10.66 & 10.79 \\
1.0 & 10.0 & 0.15 & 10.0 & 1000 & 4.87 &  4.97 \\
1.0 & 1.0 & 0.3 & 10.0 & 1000 & 10.56 & 10.67 \\
1.0 & 1.0 & 0.15 & 5.0 & 1000 & 5.91 & 5.98 \\
1.0 & 1.0 & 0.15 & 10.0 & 5000 & 18.47 & 18.96 \\
\hline
\end{tabular}
\caption{Table showing the parameters used in the various disc evolution models, and also the disc lifetimes found for each model.  The column labelled $t_1$ indicates the time at which the inner disc was drained (i.e.~when the model switches from the diffuse to direct wind profile), while the column labelled $t_2$ indicates the time at which the entire disc was dispersed.}\label{tab:disc_models}
\end{table}

\begin{figure}
\centering
        \resizebox{\hsize}{!}{
        \includegraphics[angle=270]{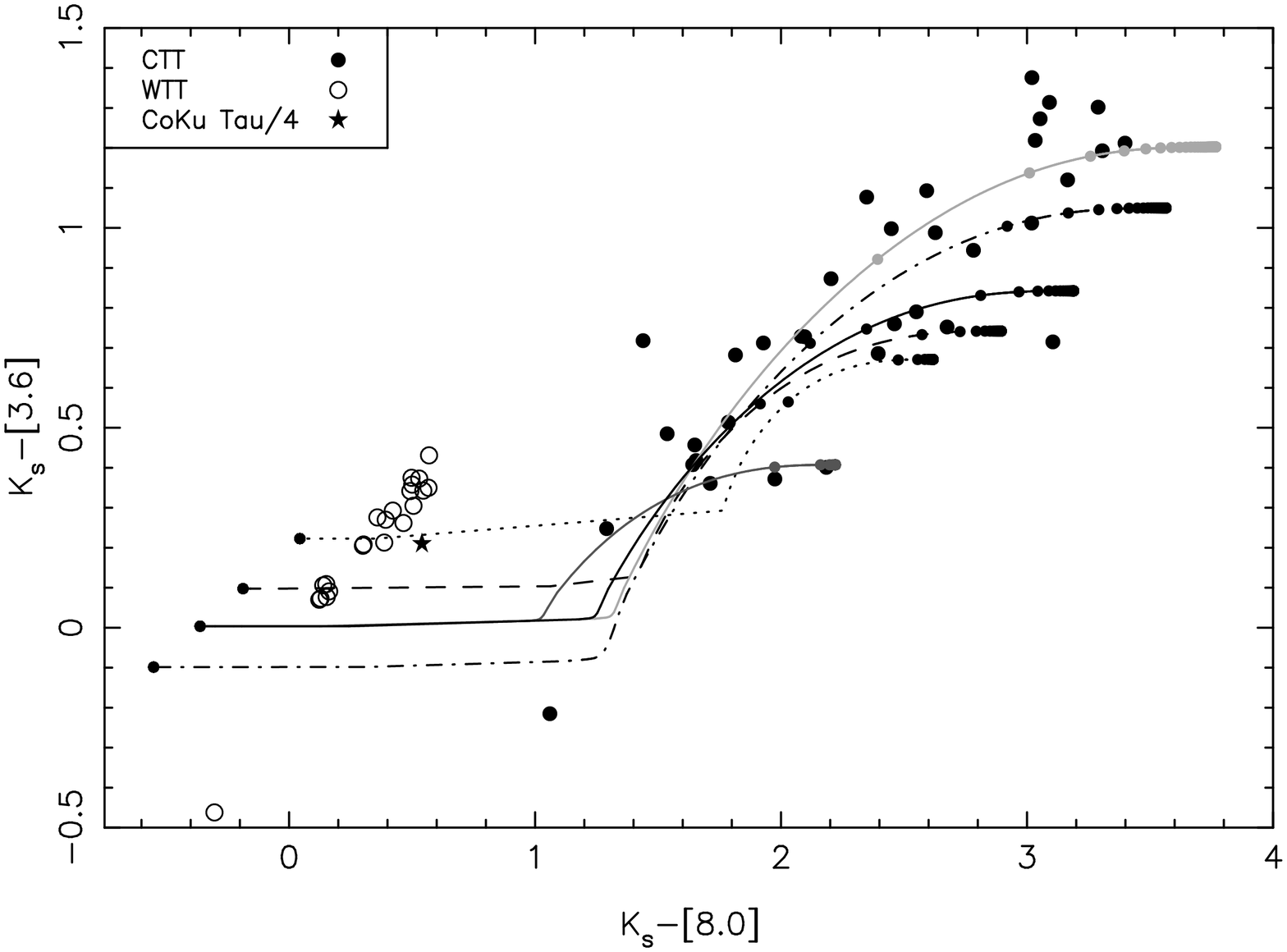}
        }

\vspace*{3mm}

        \resizebox{\hsize}{!}{
        \includegraphics[angle=270]{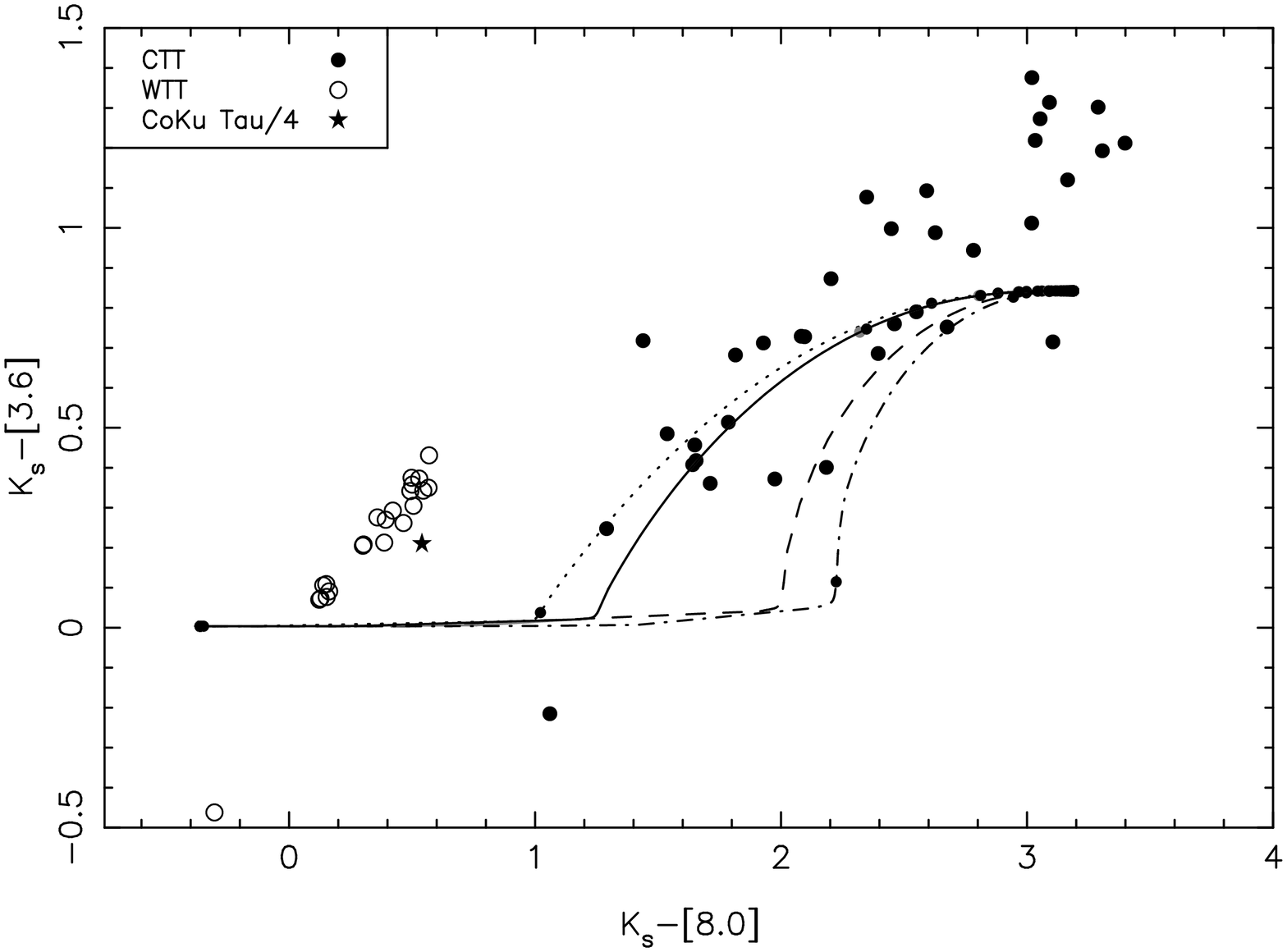}
        }
        \caption[]{2MASS/IRAC $K_{\mathrm s} - [3.6]$ / $K_{\mathrm s} - [8.0]$ plots, with data points from \citet{hartmann05}.  Solid circles represent CTTs and open circles WTTs, with the possible transition object CoKu Tau/4 represented by a star.  The upper panel shows evolutionary tracks for different stellar masses with inclination angle $i=60^{\circ}$: $M_*=0.2$ (dotted line), 0.5 (dashed), 1.0 (solid) and 2.0M$_{\odot}$ (dot-dashed).  The light grey track is for $M_*=1.0$M$_{\odot}$ with $i=0$, and the dark grey track $i=80^{\circ}$.  The lower panel shows the effect of varying the disc parameters with $M_*=1.0$M$_{\odot}$.  The solid black track is the fiducial model (as in the upper panel).  The $M_{\mathrm d}(0)=0.3$M$_{\odot}$ track is shown in grey (obscured by the black track), $\Phi=10^{43}$s$^{-1}$ as a dashed line, $R_0=5$AU as a dotted line, and $t_{\nu} = 5000 t_{\mathrm {orb}}$ as a dot-dashed line.  In both plots points are added to the tracks every $10^5$yr to illustrate the evolution.}
         \label{fig:K3K8}
\end{figure}

\begin{figure}
\centering
        \resizebox{\hsize}{!}{
        \includegraphics[angle=270]{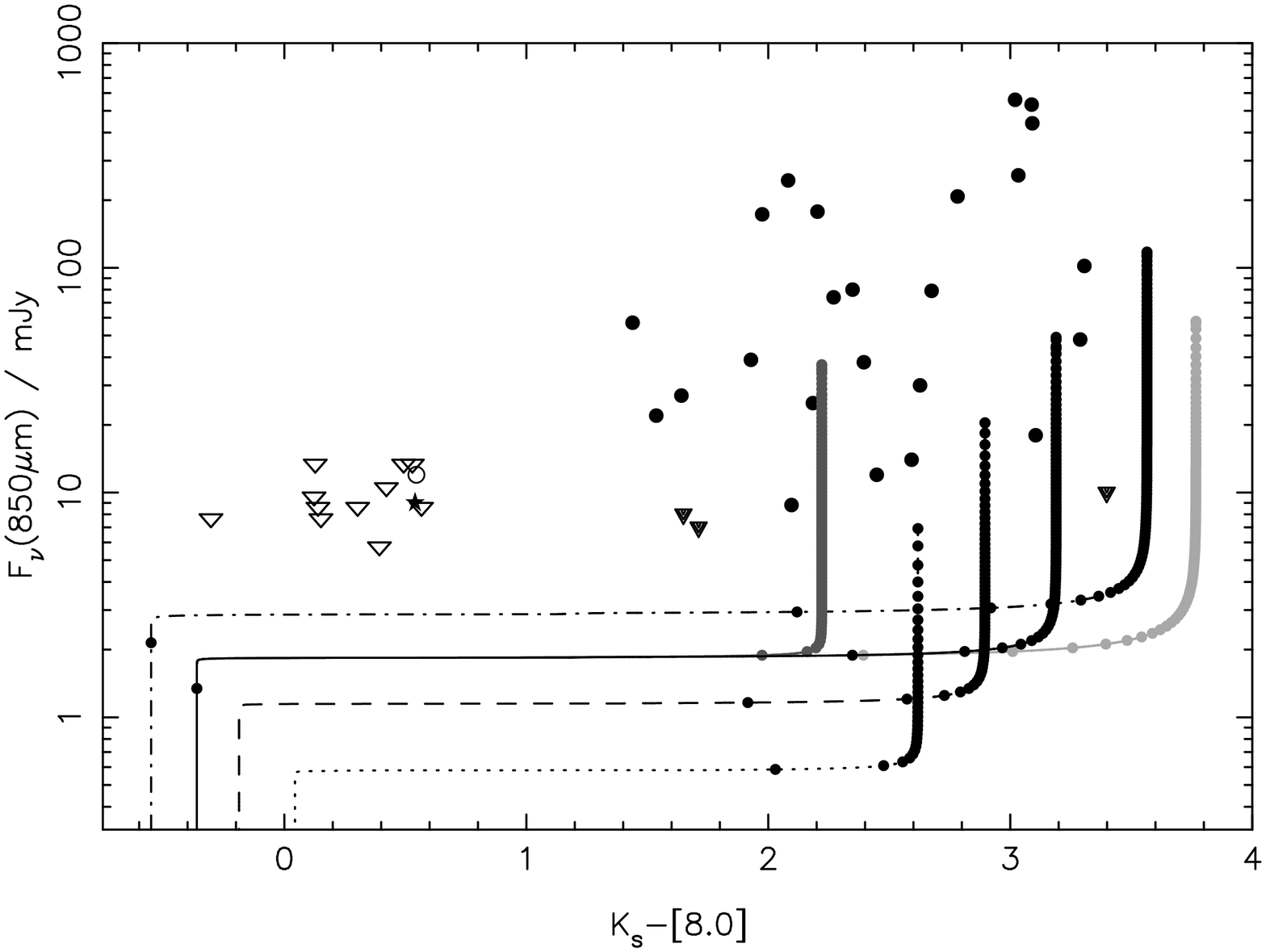}
        }

\vspace*{3mm}

        \resizebox{\hsize}{!}{
        \includegraphics[angle=270]{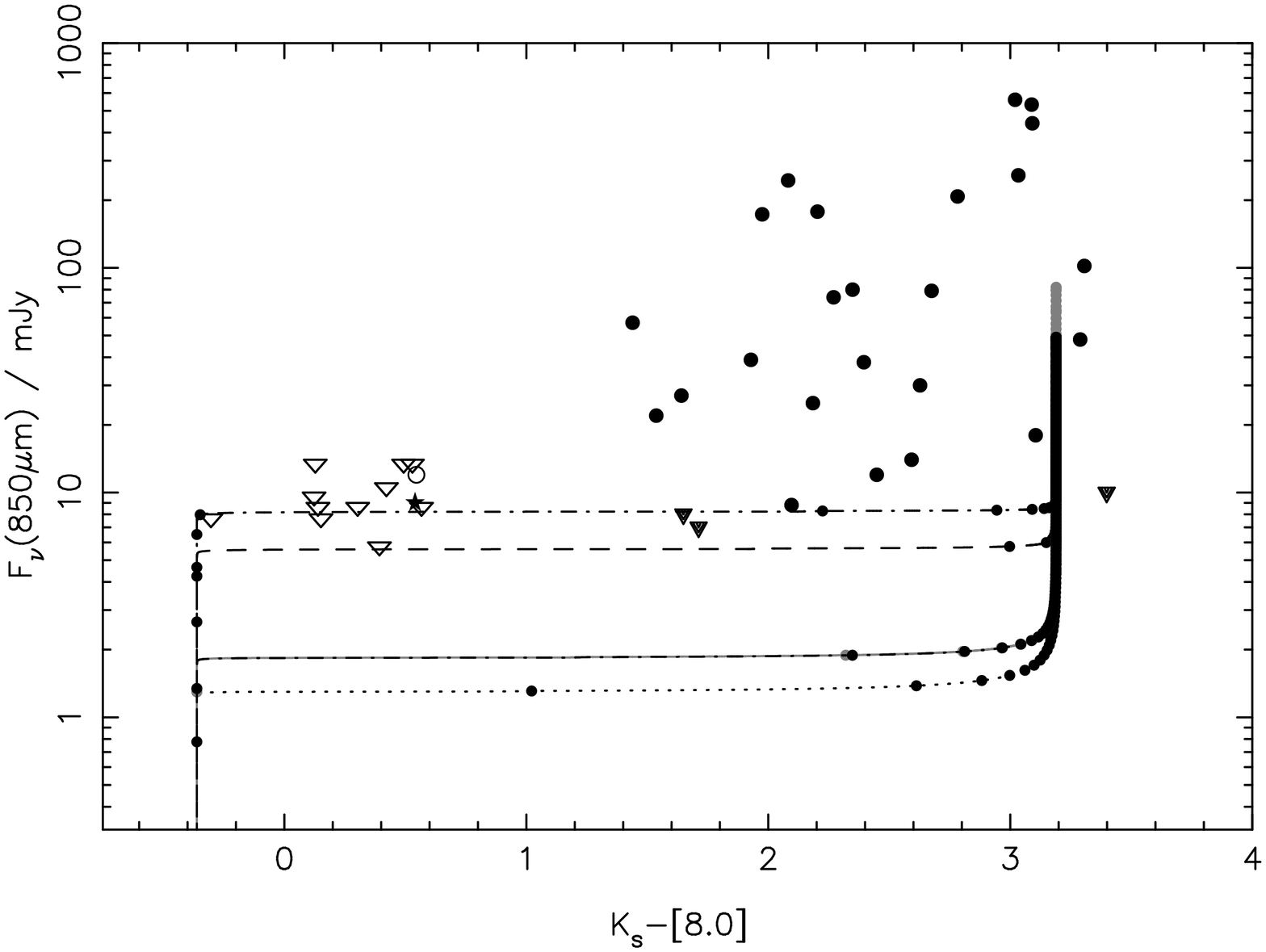}
        }
        \caption[]{$K_{\mathrm s} - [8.0]$ plotted against 850$\mu$m flux, with data points taken from \citet{hartmann05} and \citet{aw05}.  Filled symbols represent CTTs, open symbols WTTs, and the star CoKu Tau/4.  Circles represent objects detected at 850$\mu$m, while triangles denote 3$\sigma$ upper limits.  As in Fig.\ref{fig:K3K8}, the upper panel shows the effect of varying stellar mass in the models, while the lower panel shows the effect of varying the disc parameters.  The line styles are the same as those used in Fig.\ref{fig:K3K8}, with points again plotted every $10^5$yr.}
         \label{fig:SCUBA}
\end{figure}

\begin{figure}
\centering
        \resizebox{\hsize}{!}{
        \includegraphics[angle=270]{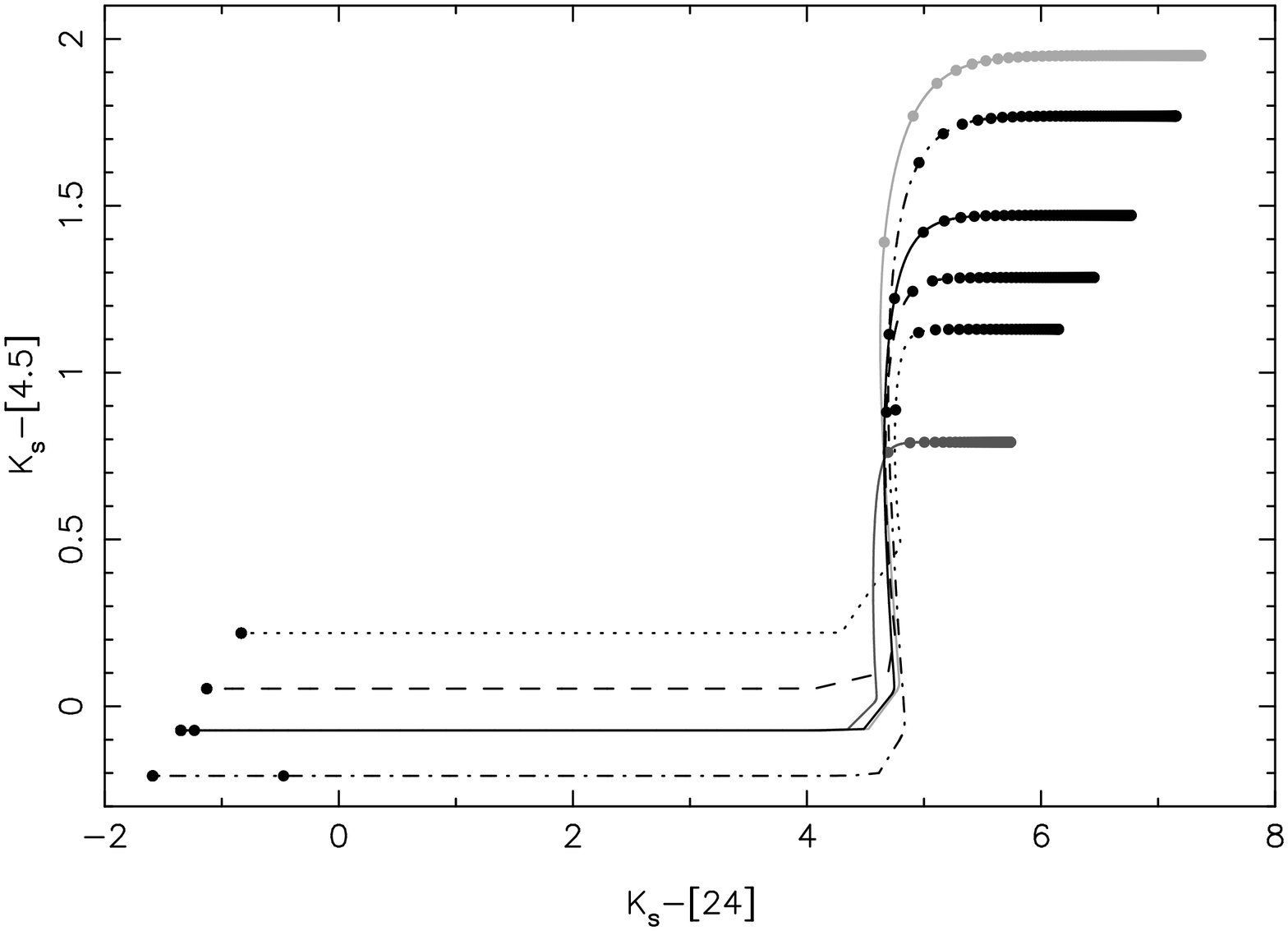}
        }

\vspace*{3mm}

        \resizebox{\hsize}{!}{
        \includegraphics[angle=270]{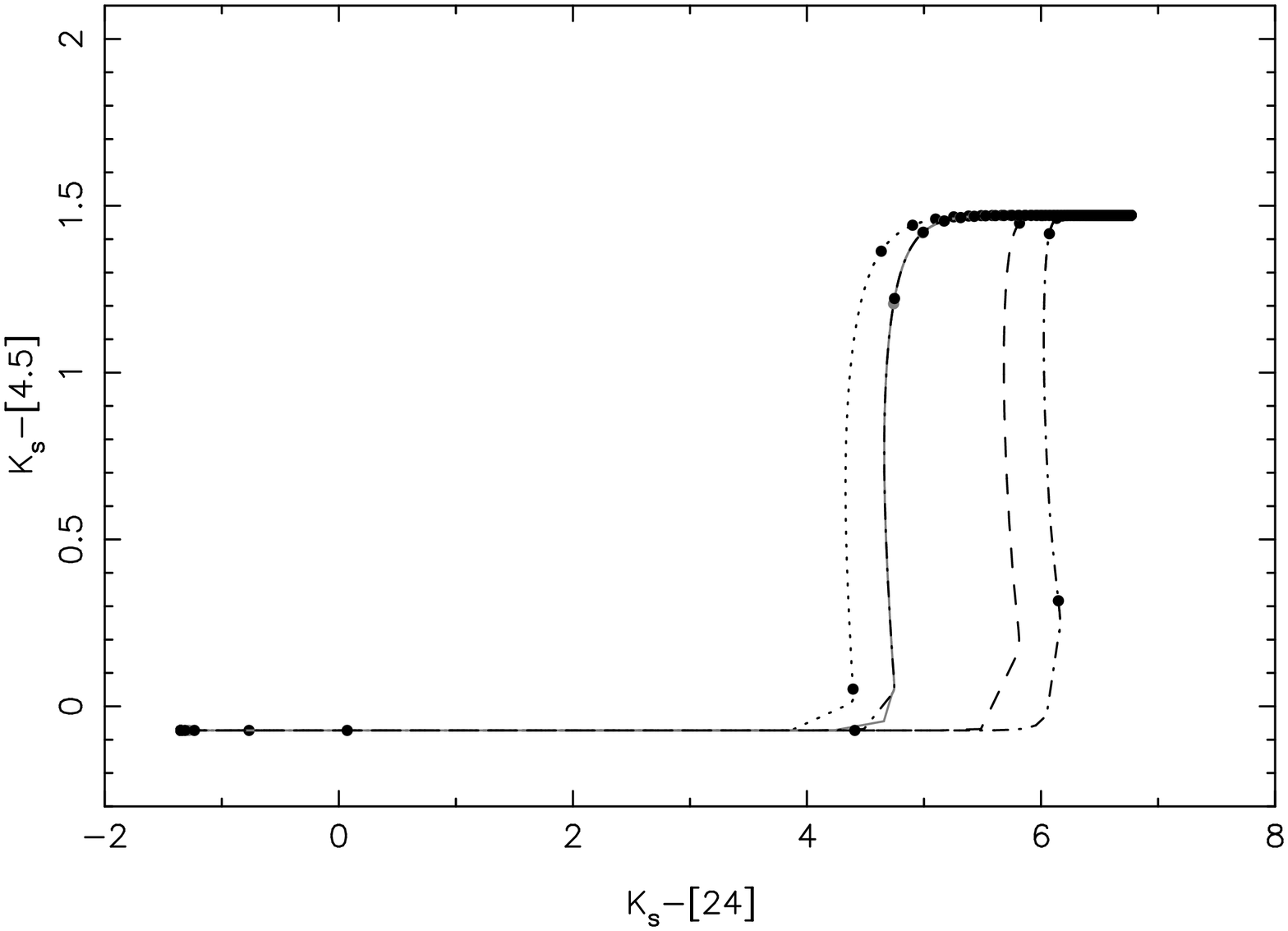}
        }
        \caption[]{Predicted evolutionary tracks in the 2MASS/IRAC/MIPS $K_{\mathrm s} - [4.5]$ / $K_{\mathrm s} - [24]$ two-colour diagram.  As in Fig.\ref{fig:K3K8} the upper panel shows the effect of varying stellar mass, while the lower panel shows the effect of varying the disc parameters.  The line styles are again the same as in Fig.\ref{fig:K3K8}, with points again plotted every $10^5$yr.}
         \label{fig:K4K24}
\end{figure}

The results of our models are shown in Table \ref{tab:disc_models} and Figs.\ref{fig:K3K8}--\ref{fig:K4K24}.  Table \ref{tab:disc_models} shows the disc lifetimes predicted by the model, which are entirely consistent with disc lifetimes of 1--10Myr and dispersal times of order $10^5$yr \citep[as derived from observations, e.g.][]{kh95,haisch01}.  Fig.\ref{fig:K3K8} shows evolutionary tracks on a $K_{\mathrm s} - [3.6]$ / $K_{\mathrm s} - [8.0]$ two-colour diagram \citep[analogous to previously published $K-L$ / $K-N$ plots, e.g.][]{kh95,act99}.  The data points show a clear gap between the loci of CTTs and WTTs which the tracks reproduce well, showing a rapid transition across the gap.  The disc emission is optically thick, and so we see Class II colours before the disc is cleared followed by Class III colours afterwards.  We also see that stellar mass and disc inclination angle are the dominant effects at these wavelengths.  For a given stellar mass and inclination angle different disc models show very similar tracks, as the infrared emission is generally optically thick.  Consequently the tracks depend primarily on the temperature profile, and are insensitive to the parameters of the disc model.  Minor optical depth effects are seen at 8$\mu$m in the models with $\Phi=10^{43}$s$^{-1}$ and $t_{\nu} = 5000 t_{\mathrm {orb}}$, but these are not significant.  The models struggle to reproduce the extreme points in the CTT distribution, at both the red and blue ends, and indeed the $K_{\mathrm s}-[8.0]$ colours of our model during the CTT phase are rather redder than the observed data.  However there are several factors which can account for this.  The SED is rather sensitive to the adopted stellar temperatures, and the disc emission can also be increased by so-called ``accretion luminosity''.  This arises due to viscous heating of the disc, and boosts the disc emission.  It is omitted from our SED model but can be significant in CTTs with high accretion rates \citep[e.g.][]{act99}.  Additionally, our model does not include emission from the inner disc edge (or ``wall'') at the dust destruction radius, which may contribute significantly to the SED at $\lambda \simeq 3$$\mu$m \citep[e.g.][]{natta01}.  We note also that our models predict rather bluer Class III colours than those observed.  This is due to the rather unrealistic use of a blackbody stellar spectrum, and we do not consider it to be a serious problem with the model.  

More interesting is the evolution of the millimetre flux, as shown in Fig.\ref{fig:SCUBA}.  The millimetre emission is mostly optically thin, and so is much more sensitive to the disc mass than the emission at shorter wavelengths.  However as it is optically thin, it is relatively insensitive to the inclination angle.  The behaviour of the tracks in the $F_{\nu}$(850$\mu$m) / $K_{\mathrm s} - [8.0]$ plane, shown in Fig.\ref{fig:SCUBA}, is explained as follows.  The initial evolution shows an increasing 850$\mu$m flux at fixed infrared colour, where the infrared colour is dependent on both stellar mass and disc inclination angle.  The increase in 850$\mu$m flux occurs because the disc viscously expands from its initial configuration, which is optically thick at 850$\mu$m.  However soon accretion takes over and the flux, now optically thin and simply proportional to the disc mass, declines at the total disc mass decreases.  The tracks fall vertically (as the infrared emission is optically thick) until the inner disc is drained, at which point the $K_{\mathrm s} - [8.0]$ colour rapidly ``jumps'' to a stellar value at a fixed millimetre flux.  The level of this flux is determined by the disc mass and temperature at this point, and does not change significantly as the inner disc drains.  This is due to the $\nu \propto R$ viscosity law, which forces most of the disc mass to reside at large radii.  Draining occurs at a fixed value of the disc accretion rate, approximately equal to that of the diffuse wind.  However the disc accretion rate can be expressed as $\dot{M} \sim \nu \Sigma$ \citep{pringle81}.  Therefore for a fixed stellar mass we expect that the level of the 850$\mu$m flux when the inner disc drains should depend only on the ionizing flux and the viscosity law.  This is verified in Fig.\ref{fig:SCUBA}, where the models with increased $\Phi$ and lower viscosity show significantly larger 850$\mu$m fluxes as the inner disc drains.  (The total disc mass at this point in the fiducial model is 0.002M$_{\odot}$.  In the model with $\Phi=10^{43}$s$^{-1}$ the disc mass at draining is 0.006M$_{\odot}$.) At this point the evolution ``stalls'' for $\sim 10^5$yr.  This occurs because most of the disc mass resides at large radii, so the 850$\mu$m flux remains approximately constant while the inner disc is cleared.  Once direct photoevaporation starts to clear the disc at large radii the 850$\mu$m flux falls rapidly to a very low level.  The original UV-switch model (CGS01), which omitted direct photoevaporation, resulted in a millimetre flux of a few mJy which remained at very late times.  This is clearly not a problem with our new model, which predicts millimetre fluxes of order $10^{-5}$mJy (the stellar contribution) once the disc has been cleared.  However the model predicts that for around $10^5$yr objects should show stellar near- to mid-infrared colours but retain millimetre fluxes of $~1$--10mJy.  This may explain the two objects detected in this region by \citet{aw05}: CoKu Tau/4 and LkH$\alpha$332 G1.  These detections are near to the sensitivity limit of current observations.  Our model predicts that a factor of $\sim10$ increase in sensitivity should results in the detection at millimetre wavelengths of an increased number of sources with Class III infrared SEDs, representing a few percent of the total population.  This represents a valuable future test of this model.

Another interesting result of the models is the predicted behaviour in the mid-infrared at 20--50$\mu$m.  This region of the spectrum probes the region outside the draining radius, but in a wavelength range where the emission is (mostly) expected to be optically thick.  To date few observations have been made here, but this should be remedied in the near future by MIPS observations.  Fig.\ref{fig:K4K24} shows the predicted tracks in the $K_{\mathrm s} - [4.5]$ / $K_{\mathrm s} - [24]$ two-colour diagram.  Here we initially see a slow decline in the 24$\mu$m emission due to weak optical depth effects.  The ``inside-out'' manner of the disc clearing causes the emission in the near infrared to decline more rapidly than that at 24$\mu$m, with the models predicting that a significant population of sources, again at the few percent level, should show Class III colours in the IRAC bands but significant excesses at 24$\mu$m.  Forthcoming observations from {\it Spitzer} will therefore provide another valuable test of the model.

Lastly, we note that it is possible to constrain some disc parameters from the model results.  By demanding that disc lifetimes be in the 1--10Myr range, as seen in observations \citep[e.g.][]{haisch01}, it is possible to constrain the disc viscosity, albeit rather weakly.  For initial disc masses of 0.15$M_*$ the ratio of viscous to orbital times must be $\simeq 1000$, and certainly less than 5000.  For realistic $H/R$ ratios this suggests a viscosity parameter in the range $0.02 \lesssim \alpha \lesssim 0.4$.  There is no real consensus as to what value of $\alpha$ can be produced by models of angular momentum transport in discs, but fiducial values tend to be of order $\alpha\sim0.01$ \citep{stone_ppiv}.  Thus these derived values of $\alpha$ are near to the upper limit of those predicted by current models of angular momentum transport in discs, and may pose problems for these models.  

Similarly, by comparing the models to observed millimetre fluxes (as in Fig.\ref{fig:SCUBA}) it is possible to place indirect constraints on the ionizing flux $\Phi$.  The fact that few CTTs are not detected at 850$\mu$m at the 10mJy level, and that similarly few WTTs are detected at the same level, suggests that the millimetre flux during the ``transition'' phase must be in the 1--10mJy range for most objects.  (If this were not the case we would expect to see either many more WTTs detected or many more CTTs not detected at the 10mJy level at 850$\mu$m.)  As seen in Fig.\ref{fig:SCUBA} this level depends on both the disc and stellar parameters, and not solely on $\Phi$.  However the results suggest ionizing fluxes in the range $\sim 10^{41}$--$10^{43}$s$^{-1}$, a range consistent with our previous estimates \citep{chrom}.

\section{Discussion}\label{sec:dis}
There are obvious limitations to the model.  Firstly, we note that the manner in which we treat the disc temperature and viscosity profiles are not self-consistent.  The derived wind profiles (from Paper I) assume a constant $H/R$, but throughout this paper we adopt a disc temperature profile which implies disc flaring.  Further, although the temperature profile adopted when evaluating the SED does result in a $\nu \propto R$ viscosity law, the normalisation of the two parametrizations is inconsistent.  However the viscosity depends primarily on the midplane temperature, while the emitted SED depends on the surface temperature.  These are most likely not the same \citep[e.g.][]{cg97}, and so such an inconsistency is not unreasonable.  Therefore while there are minor inconsistencies in the disc model, they do have some physical motivation and they do not have a strong effect on the model results.

In addition the SED model is rather simple, employing a power-law for the disc temperature structure and neglecting several possibly important factors, such as viscous heating or emission from the inner wall.  Further, the temperature normalisation is rather sensitive to the stellar temperature adopted, as seen in Equation \ref{eq:temp_norm}.  The flux emitted by the disc is very sensitive to disc temperature, so the output SEDs are in turn rather sensitive to both the stellar temperature and the normalisation condition.  Our models struggle to reproduce the reddest and bluest CTTs in the data and, as seen in Fig\ref{fig:K3K8}, our simple model predicts rather redder $K_{\mathrm s}-[8.0]$ colours than those observed during the CTT phase.  However this is easily remedied by small alterations to the SED model, and we note that our simple treatment of the disc neglects several possibly important effects (see Section \ref{sec:results}), such as accretion luminosity or emission from the inner disc wall.  Additionally, our model uses a single temperature blackbody to model the stellar flux.  Model atmospheres can differ markedly from blackbody spectra in the near-infrared \citep[e.g.][]{baraffe98}, and this may also affect the colours obtained from the SED model.  The model also struggles to reproduce the handful of objects with the largest observed 850$\mu$m fluxes.  However, as seen above, the emission at millimetre wavelengths is extremely sensitive to the disc parameters.  Additionally, these objects tend to have rather unusual SED slopes in the sub-millimetre \citep{aw05}, and so we do not consider this to be a serious problem with the model.

A further consideration is the manner in which the transition from the diffuse to direct regime is treated (see Section \ref{sec:disc_model}).  We define a critical value for the optical depth to ionizing photons along the disc midplane and use it to switch instantaneously between the two wind parametrizations.  A more realistic treatment would gradually increase the strength of the direct field as the inner disc is drained.  Additionally, the behaviour of the diffuse field as the inner disc drains should be considered in more detail, as the diffuse wind model assumes that the underlying disc is always optically thick to ionizing photons.  The transition occurs very rapidly, and its treatment does not have a strong effect on the overall evolution of the disc.  The transition is important, however, when we consider the detailed conseqences for the observed SED during clearing.  This is because the inner disc becomes optically thin to infrared emission (perpendicular to the disc midplane) rather earlier than it becomes optically thin to Lyman continuum photons (along the midplane).  Objects appear as ``inner hole'' sources during this phase, and the wavelength at which the disc emission ``cuts in'' is rather sensitive to the SED model.  

Recently \citet{mccabe05} have observed a strong correlation in passive (non-accreting) discs, between stellar spectral type and the wavelength at which outer disc emission becomes visible.  They  define ``passive discs'' as sources which show excess emission at 10--12$\mu$m but lack any near-infrared excess, and in a survey of binary secondary stars they find that that stars of later spectral type are more likely to possess passive discs than stars of earlier spectral type.  The binary separations are large, so within the framework of a photoevaporation model they attribute this to the linear scaling of the draining radius with stellar mass.  Thus low mass stars have smaller holes, which they argue should contribute disc emission at shorter wavelengths than the corresponding holes in higher mass stars. This interpretation is predicated on the assumption that the smaller holes are hotter.  However, in our models, smaller holes turn out to be only modestly hotter, since the stellar photospheric temperature (which sets the temperature of the reprocessing disc) is lower for less massive stars.   Further, this interpretation remains valid only if the inner disc clearing takes significantly longer than the outer disc clearing.  If the outer disc clearing dominates the dispersal time then the distribution of observed hole sizes at any given time is independent of stellar mass.  Therefore a more accurate treatment of the transition from the diffuse to direct wind is necessary in order to make such detailed predictions.  We also stress that a better treatment of the disc SED, in particular the contribution from the inner wall, is needed before we can assess whether our models can reproduce the passive discs observed by \citet{mccabe05}.

We note also that our model neglects the effects of non-ionizing far-ultraviolet (FUV) radiation on the evolution of the disc.  FUV radiation heats the disc to lower temperatures than ionizing radiation \citep[$\simeq 1000$K cf.~$\simeq 10,000$K, ][]{jhb98,gh04} and therefore FUV-driven winds are launched from rather larger radii than those driven by ionizing radiation, typically launching at radii from 20--100AU.  \citet{adams04} have shown that external FUV radiation can produce a significant disc wind, which is is dominated by flow from the outer disc edge.  However this wind relies on the presence of a strong interstellar FUV radiation field, which is only present in the immediate vicinity of massive O- or B-type stars.  Self-consistent models of FUV heating from the central object have not yet been completed, but it seems likely that FUV radiation can drive a wind from the outer disc edge in this case also.  The interplay between mass-loss from the outer disc edge (from FUV heating), mass-loss from smaller radii (from Lyman continuum heating) and viscous evolution will provide an interesting area for future study.  While the details of such a calculation are beyond the scope of this paper, we note that if the stellar FUV radiation field can drive a significant wind from the outer edge of the disc then this may significantly shorten the disc lifetime in any given disc model.  Consequently we expect that the inclusion of such a wind will reduce the viscosity ($\alpha$) required to satisfy observational constraints on disc lifetimes.  

Another important consideration is dust.  Dust is responsible for all of the emission discussed above, and our SED modelling assumes a constant dust-to-gas ratio throughout.  However it is not at all obvious that the dust in the disc remains coupled to the gas.  There are several forces which act on dust grains in discs, all of which may be important here \citep[see e.g.][]{gustafson94}.  In addition to gravity the dust grains feel a drag force from the gas, and grains can both grow and be ground down through collisional processes.  Grains can also feel a force due to radiation pressure.  In most cases the optical depth through the disc is such that this force is negligible, but near to the inner disc edge it may become significant.  Radiation from the star is also responsible for the Poynting-Roberston (P-R) effect, which can cause grains to lose angular momentum and slowly spiral inwards.  Again, this effect is not significant if the grains are shielded from the stellar radiation field, but may become significant once the gas disc begins to drain.  Additionally, as noted in Paper I, any dust remaining in the ``inner hole'' may absorb some ionizing photons, reducing the efficiency of the wind and lengthening the dispersal time-scale.

The net effect of these forces on the dust in the disc is difficult to predict without making detailed calculations.  The evolution of dust in the original UV-switch model has been studied in some detail by \citet{tcl05}.  They find that gas drag causes millimetre-size dust grains to accrete on to the star more rapidly than the gas unless the grains are rather ``fluffy'' (i.e.~have rather low densities).  In this model the grains are removed from the outer disc before the inner disc is drained, and the decline in disc emission is primarily due to the rapid accretion of the grains rather than global evolution of the disc.  However the sub-millimetre observations by \citet{aw05} do not detect a significant fraction of objects which have no dust in the outer disc but are still accreting gas from their inner disc, suggesting that the dust is removed too quickly in the model of \citet{tcl05}.  Additionally, \citet{tcl05} find that the force exerted on the grains by the wind is always less than gravity, so little dust is carried away by the photoevaporative wind.  This is not of great significance in their model, as the grains are removed from the outer disc by gas drag, but may be a problem when considering direct photoevaporation of the outer disc.  In this case gas drag cannot migrate the grains beyond the inner disc edge, and the wind cannot carry away any dust grains found in the outer disc.  Consequently we would expect the dust grains in the outer disc to remain after the gas has been dispersed.  However the time-scales for grains to be removed by either collisional processes or by P-R drag are very short, of order $10^3$--$10^4$yr at radii of a few AU \citep[e.g.][]{gustafson94}.  Consequently any dust ``left behind'' in this manner is expected to be removed sufficiently quickly as to have a negligible effect on the evolution of the SED.

However the analysis of \citet{tcl05} does not consider the effects of radiation-powered forces on the grains.  This is not a significant problem in their model as the dust is shielded from stellar irradiation, but may well become significant in our model once the inner disc begins to drain.  \citet{kl01,kl05} found that radiation pressure in discs with low opacity can result in a clumping instability in the dust grains.  In our model this effect may well become important when the inner disc drains, as the optical depth is essentially zero inside the inner disc edge.  Such clumping increases grain collision rates, and therefore may have important consequences for planet formation theories as well as on the observed SED.  We also note that any dust exposed to the stellar radiation field in this manner will attenuate the direct radiation field responsible for the photoevaporative wind (see discussion in Paper I).  Additional effects, such as grain growth and dust replenishment, may also be significant.  In short, it is unlikely that the disc evolves with a constant gas-to-dust ratio, as we have assumed.  However the manner in which the dust evolves is not at all obvious, and we make no attempt to model it here.   We also note in passing that a handful of gas-poor discs have recently been observed by \cite{nw05}, and suggest that photoevaporation may have played a significant role in their evolution.

\subsection{Inner holes}
Lastly, we note that one consequence of the model is that all objects are predicted to pass through an ``inner hole'' phase, where the disc is drained close to the star but remains ``normal'' at larger radii.  Several such objects have now been observed (e.g. GM Aur, \citealt{rice03}; DM Tau, \citealt{calvet05}; TW Hya, \citealt{calvet02}; CoKu Tau/4, \citealt{forrest04}), and it has been suggested that these may represent a class of ``transition objects'', intermediate between the CTT and WTT states.  It is not clear that these objects represent a homogeneous group (in fact this seems unlikely), but it is interesting to compare them to the predictions of our model.  

Despite the presence of holes in their (dust) discs, both GM Aur and DM Tau have large accretion rates (on to the star), of order $10^{-8}$M$_{\odot}$yr$^{-1}$ \citep{calvet05,hg05}.  Additionally, in the case of GM Aur the observed inner hole is incompatible with a standard gas-to-dust ratio \citep{rice03}.  As such GM Aur is obviously inconsistent with the photoevaporation model, and some other process, such as grain growth or the presence of a planet, must be at work here.  Both GM Aur and DM Tau have also been observed to have relatively large outer disc masses, of order 0.05--0.1M$_{\odot}$ \citep{calvet05,hg05}.  In the photoevaporation model the disc mass at draining is approximately equal to the product of the (diffuse) wind mass-loss rate ($\sim 10^{-10}$M$_{\odot}$yr$^{-1}$) and the viscous evolution timescale ($\sim 10^7$yr), and, as seen in Section \ref{sec:ref_results}, is typically 0.001M$_{\odot}$ (plus or minus around an order of magnitude).  Therefore it is rather difficult (though not completely impossible) for photoevaporation to produce inner holes with outer disc masses as large as those observed in these two sources, again suggesting that photoevaporation is not responsible for the holes observed in these two discs.  However we note that in deriving these disc masses \citet{calvet05} have derived rather small viscosity parameters ($\alpha \sim 0.001$), and that larger values of $\alpha$ would imply correspondingly smaller disc masses \citep[see also][]{hg05}.  

TW Hya has a ``hole'' in the disc inside 3--4AU, but still shows some dust and gas emission from inside this radius \citep{calvet02}.  It also has a measured accretion rate of $\simeq 4\times 10^{-10}$M$_{\odot}$yr$^{-1}$ \citep{muz00}, and it has been suggested that a planet may have formed (or be forming) in the disc.  The disc mass is rather larger than that predicted by the photoevaporation model, but the low accretion rate is similar to that produced as the inner disc drains in the photoevaporation model.  Therefore in principle it may be possible to model the SED with a photoevaporation model, although in order to reproduce the observed accretion rate the model requires that the object be observed during a rather short ``window'' in the evolution (while the inner disc is in the process of draining).  Consequently it is not yet clear whether or not photoevaporation plays a role here.  

CoKu Tau/4, however, is almost entirely devoid of material inside 10AU and shows no evidence of accretion on to the star \citep{forrest04,dalessio05}.  The outer disc mass of 0.001M$_{\odot}$ \citep{dalessio05} is similar to that predicted by our model (see Section \ref{sec:ref_results} and discussion above).  We have seen above that the evolutionary tracks produced by the photoevaporation model can reproduce the SED of CoKu Tau/4 well, and at present this object seems to be entirely consistent with the predictions of the photoevaporation model.  

We note that much current work has invoked the presence of a formed or forming planet in order to explain such holes \citep[e.g.][]{calvet02,rice03,quillen04}.  Whilst this may well turn out to be the case, we emphasise that other mechanisms, such as photoevaporation, can also produce holes in discs, and that holes do not necessarily imply the presence of a planet.  Indeed it seems likely that photoevaporation and planet formation occur simultaneously during the evolution of protoplanetary discs, and it is not yet clear whether photoevaporation acts to accelerate or retard planet formation.
{\it Spitzer} is expected to discover many more objects similar to those discussed here, and only with more data will the nature of these objects become apparent.  However the current data suggest that the inner hole sources discovered to date are far from a homogeneous class of objects.  


\section{Summary}\label{sec:summary}
We have presented a new model for the evolution of protoplanetary discs.  Our model combines viscous evolution with photoevaporation of the disc by stellar radiation, and includes the effects of the direct radiation field at late stages of the evolution (using the wind prescription derived in Paper I).  We have constructed a numerical model for the evolving disc, and used a simple prescription to model the behaviour of the spectral energy distribution as the disc evolves.  Our fiducial model predicts that the disc is completely dispersed on a time-scale of order $10^5$yr after a lifetime of a few Myr, consistent with observationally derived time-scales.  Our results are consistent with observational data across a broad range in wavelengths, and allow us to place weak constraints on some model parameters: we derive ionizing fluxes in the range $\sim 10^{41}$--$10^{43}$s$^{-1}$ and viscosity parameters ($\alpha$) in the range $\sim 0.02$--0.4.  We also make predictions as to what will be seen in future observations in the mid-infrared and at millimetre wavelengths.  To date this is the only model of disc evolution which can reproduce the rapid disc dispersal seen in observations of T Tauri discs.  We note also that the model suggests that all evolving discs pass through a short ``inner hole'' phase.  During this phase the outer disc mass is of order 0.001M$_{\odot}$, and we predict that inner hole sources should represent around 1--10\% of the observed population of TTs.  We compare the predictions of the model with the handful of inner hole sources observed to date and show that some, but not all,  of these objects are consistent with the predictions of our model.

\section*{Acknowledgements}
We thank Doug Johnstone, Andreea Font and Ian McCarthy for providing numerical results from their hydrodynamic models.  We also thank Doug Johnstone for a useful referee's report, which improved the clarity of the paper.  We thank Chris Tout for providing data from his pre-main-sequence evolutionary models.  RDA acknowledges the support of a PPARC PhD studentship.  CJC gratefully acknowledges support from the Leverhulme Trust in the form of a Philip Leverhulme Prize.  Parts of this work were supported by NASA under grant NNG05GI92G from the Beyond Einstein Foundation Science Program.


\label{lastpage}

\end{document}